\documentclass[amssymb,amsmath,pra,twocolumn,floatfix,showpacs,superscriptaddress]{revtex4-1}
\usepackage{graphicx}
\usepackage{lastpage}
\usepackage[normalem]{ulem}
\usepackage[utf8]{inputenc} 
\usepackage{IEEEtrantools}
\usepackage{listings}
\usepackage{amsmath}
\graphicspath{ {Images/} }
\usepackage{epstopdf}

\begin{document}
\title{Geometrogenesis under Quantum Graphity: problems with the ripening Universe}

\author{Samuel A.~Wilkinson}
\affiliation{Chemical and Quantum Physics, School of Applied Sciences, RMIT University, Melbourne 3001, Australia}
\email{samuel.wilkinson@rmit.edu.au}

\author{Andrew D.~Greentree}
\affiliation{Chemical and Quantum Physics, School of Applied Sciences, RMIT University, Melbourne 3001, Australia}
\affiliation{ARC Centre of Excellence for Nanoscale BioPhotonics }

\date{\today}

\begin{abstract}
Quantum Graphity (QG) is a model of emergent geometry in which space is represented by a dynamical graph. The graph evolves under the action of a Hamiltonian from a high-energy pre-geometric state to a low-energy state in which geometry emerges as a coarse-grained effective property of space. Here we show the results of numerical modelling of the evolution of the QG Hamiltonian, a process we term ``ripening" by analogy with crystallographic growth. We find that the model as originally presented favours a graph composed of small disjoint subgraphs. Such a disconnected space is a poor representation of our universe. A new term is introduced to the original QG Hamiltonian, which we call the \textit{hypervalence} term. It is shown that the inclusion of a hypervalence term causes a connected lattice-like graph to be favoured over small isolated subgraphs.

\end{abstract}


\maketitle

\section{Introduction}
The idea that space-time geometry may not be a fundamental feature of the universe has arisen in several speculative models for the microscopic structure of space \cite{bib:Sind2012}, in particular those which aim to formulate a background independent theory of quantum gravity \cite{bib:Mark2009}. This approach is taken in research paradigms such as causal dynamical triangulation \cite{bib:AJL2004}, group field theory \cite{bib:Oriti2007} and structurally dynamic cellular networks \cite{bib:Req2015}. It has been suggested that the AdS/CFT correspondence in string theory also indicates that space-time is not fundamental \cite{bib:Seib2006}. To consider geometry to be merely an effective, coarse-grained property of space-time may resolve some of the conflicts between quantum theory and general relativity \cite{bib:Oriti2013}. However, this comes at great technical and conceptual cost. A quantum theory with no inherent notion of geometry cannot be formulated with respect to a fixed manifold, nor can it make use of the usual global symmetries. For this reason, models of emergent geometry often use the tools of combinatorics.

One such model is Quantum Graphity (QG) \cite{bib:KMS2008}, which postulates that space is represented as a dynamical graph, where points in space are represented by vertices of the graph and adjacencies in space are represented by edges in the graph. Since QG uses simple, undirected, unembedded graphs, there is no inherent notion of geometry at the fundamental level in this picture. Rather, it is postulated that this model will exhibit two thermodynamic phases: a high temperature phase in which there is no sensible notion of geometry, and a low temperature phase in which geometry emerges as an effective phenomenon. The very early universe is believed to have been in the high temperature phase, undergoing a phase transition at some later time to the low temperature phase. This process is called \textit{geometrogenesis}. Since we are interested in examining an emergent property of the low-temperature condensed phase, it is natural to draw parallels with condensed matter physics, and apply intuitions developed there to the problem at hand. In this way, QG can be viewed as a member of the newly developing family of condensed matter analogue models of spacetime, which treat space or spacetime itself as though it were a many-body system \cite{bib:Bain2013}.

The interpretation of temperature in QG is not clear. If geometry is considered to be a collective phenomenon, and we are to appeal to statistical physics to explain it, then the role of temperature will be significant. However, when the system under investigation is space-time itself, temperature is difficult to define. In this work we interpret matter and energy fields living in space-time as a heat bath. The graph can lower its energy by creating matter. This process is not considered in detail here, however the string-net condensation of Levin and Wen \cite{bib:PhotEmerg, bib:Wen2005} has been proposed as a mechanism by which matter may be emergent from a QG space-time \cite{bib:KMS2008}.

The low temperature phase of QG is assumed to be a lattice graph, with the geometry of the lattice corresponding to the emergent geometry of space-time \cite{bib:KMS2008}. In particular, parameters of the model are often chosen so that a regular 2-dimensional honeycomb graph is favoured as the ground state. This is clearly not a representation of our 3-dimensional spatial geometry, but is a useful simplification for studying some general features of the model. The high-energy pre-geometric state has usually been considered to be the complete graph, in which every vertex is connected to every other. This graph has no geometrical interpretation, making it a candidate for pre-geometry. Furthermore, thermodynamic arguments support the notion that higher temperature graphs have more edges, making the graph with the most edges possible a natural maximum temperature limit \cite{bib:Cara2011}. However, the empty graph, in which there exist vertices but no edges, so that every point in space is completely isolated, is also a candidate high-temperature pre-geometric graph \cite{bib:Wilk2014}. This is perhaps more in line with the concept of a universe arising spontaneously from ``nothing", and offers a stronger concept of ``spacelessness".

The geometrogenic phase transition in this model is of great significance and has been investigated from a number of perspectives. When recast on a line graph, the model can be reduced to an analogue of an Ising model, making it amenable to a mean-field treatment. In this approach, using the average degree of the graph as an order parameter, it was found that as the number of vertices goes to infinity, the critical temperature of the geometrogenic phase transition goes to zero \cite{bib:Cara2011}. Another approach has been to estimate the partition function of the model by considering which types of graphs will contribute most \cite{bib:Kono2008}. The two main competing families of graphs are homogeneous lattice-like graphs (due to their low energy and consequent high Boltzmann factor) and random graphs (despite being higher in energy, the large number of random graphs means that they contribute significantly). Phase transitions between the two states (lattice-dominated and random-dominated) are found once again to occur at zero temperature in the $N \rightarrow \infty$ limit. This behaviour is also seen in other related graph models which exhibit geometrogenesis \cite{bib:ChenPlot2013, bib:Conr2011}. A zero-temperature phase transition may imply that the phase transition does not exist, indicating that the model is faulty. Alternatively, it could mean that geometrogenesis is a \textit{quantum} phase transition, rather than a \textit{thermal} phase transition, and must be modelled accordingly. Since its description in QG, the event of geometrogenesis has been shown to have a concrete realisation in causal dynamical triangulations, and may also be present in loop quantum cosmology \cite{bib:Miel2014}.

In this work we study the growth (ripening) of grains from the empty graph under the epitaxial approximation (discussed below). We have assumed that the formation of spatial grains from a pre-geometric graph follows a process analogous to the growth of crystals. Instead of a homogeneous transition across the entire graph, we argue that as the graph lowers its energy it will tend to form distinct grains which then knit together to form a geometric space. Over time the size of these grains may increase until they are sufficient to house the Universe. This picture for the evolution of the graph lends itself to the formation of domain boundary defects, which have been shown to have observable consequences \cite{bib:QSM+2012}.

Numerical simulations of ripening under the epitaxial approximation reveal a tendency for the model to form a graph consisting of small isolated subgraphs in the low-energy regime. Evolution from the empty graph under the epitaxial approximation leads to a disjoint graph as the lowest energy state of the system, and we show that such disjoint graphs are inevitably lower in energy than connected graphs, given the Hamiltonian and parameters first proposed for this model \cite{bib:KMS2008}. The formation of a disjoint low-energy graph was first noted in \cite{bib:Wilk2014}, and here we demonstrate that it remains even under more accurate calculations. A disjoint ground state is clearly undesirable for a model that aims to reproduce an extended geometric space. Therefore, we propose a modification to the model which may lead to the formation of a connected space.

\section{Model}

We begin by postulating that space may be represented by a simple, undirected graph $\mathcal{G}$, which is defined by two sets $\mathbb{V}$ and $\mathbb{E}$. $\mathbb{V}$ is the set of \textit{vertices} (or nodes) on the graph, and $\mathbb{E}$ is the set of \textit{edges}, which are unordered pairs of the elements of $\mathbb{V}$. Two vertices in $\mathbb{V}$, $\nu_i$ and $\nu_j$, are said to be connected (or adjacent) if there exists in $\mathbb{E}$ an edge $(\nu_i,\nu_j)$. This graph is dynamic, in the sense that the set of edges may change in time, however the set of vertices is taken to be fixed.

The energetics of the graph are determined by the Hamiltonian, 
\begin{align}
H = H_V + H_L + H_{\textrm{hop}}.
\label{eq:Ham}
\end{align}
The valence term $H_V$ assigns an energy to the graph based on the valence, or degree, of each vertex,
\begin{align}
H_V = g_V \sum_{i} e^{p(v_i - v_0)^2}
\label{eq:Val}
\end{align}
where $v_i$ is the valence of the $i^{th}$ vertex, $g_V$ is a positive coupling constant, $p$ is a dimensionless real number and $v_0$ is the ``ideal" valence of the graph. $H_V$ is minimised when every vertex in $\mathcal{G}$ has $v_i = v_0$.

The loops term $H_L$ reduces the energy of the graph when there are more loops,
\begin{align}
H_L = -g_L \sum_{L = 3}^{L_{\textrm{max}}} \frac{r^L}{L!} \sum_i P(L,\nu_i)
\label{eq:loops}
\end{align}
where $P(\nu_i,L)$ is a function that counts the number of loops of length $L$ that pass through vertex $\nu_i$, $r$ is a dimensionless real number and $g_L$ is a positive coupling constant. The sum over loop lengths $L$ begins at 3, because that is the length of the shortest possible non-retracting loop. Ideally, the sum would extend upwards to include loops of infinite length, but to the make the model computationally tractable, loop counting is truncated at some maximum length $L_{\max}$. The weighting factor $r^L/L!$ is small both when $L$ is small and when $L$ is large. Between these points it reaches a peak at some value $L^*$, which is determined by $r$. Thus arbitrarily long loops contribute a negligible amount of energy, justifying the use of a truncation length $L_{\max}$, and by varying $r$ we can tune the Hamiltonian so that loops of some desired length $L^*$ contribute most. As $r$ is varied, we find that 5-loops are dominant over 6-loops for values of $r$ less than 6, and that 7-loops are dominant over 6-loops for values of $r$ greater than 7. Between these points, $L^* = 6$, so that 6-loops contribute most significantly to $H_L$. Therefore, we have chosen a value for $r$ in the middle of this region, $r=6.5$. This differs from the value of $r=7.1$ which was used in previous literature \cite{bib:KMS2008}. Repeating our calculations with each of these two values of $r$ shows no significant difference to our results except where stated.

The negative sign in $H_L$ means that this term lowers the energy of the graph, so that it favours graphs with many loops of length $L^*$. The loop counting function $P(a,L)$ is determined so that each unique loop is counted only once. If such a procedure were not used, then a symmetry factor of $1/(2L)$ would be required to account for non-unique loops.

Rather than count all loops in the graph, we only consider shortest-path (SP) loops, as these are the most relevant for characterizing a lattice-like graph with emergent geometry. SP loops are those that contain no ``short-cuts", so that the distance between two vertices along the loop is equal to their distance on the graph. While there exist explicit algebraic formulas for counting the number of loops on a graph up to loops of length 7, there exist no such formulae for SP loops. SP loops are therefore calculated algorithmically using the method presented in \cite{bib:Franzblau}.

Finally, $H_\textrm{hop}$ is a kinetic term that allows edges to propagate through the graph, changing the configuration of the graph. Here we are not concerned with the particular form of this term, but include it so that the configuration of the graph may evolve in time. In general, adding, deleting or hopping edges does not conserve energy. This can be interpreted as the graph being in contact with some heat bath, although when our graph represents space itself the idea of an external heat bath is problematic \cite{bib:Cara2011}. We follow previous work \cite{bib:Hamma2010} and take the heat bath to represent the creation and annihilation of matter on the graph, so that the lowing of the energy of the graph can be interpreted as the creation of matter and the total energy of the graph + matter system is conserved. Since we neglect matter in this work, the evolution of the graph is a non-unitary process.

As discussed in the Introduction, we take the empty graph as the initial state of the system.

\section{Ripening Under the Epitaxial Approximation} \label{sec:RipeEpitaxy}

The epitaxial approximation, introduced in \cite{bib:Wilk2014}, assumes that evolution of the graph happens at only one edge at a time. First, we take the state of the graph to be fixed. Then, we add or delete a single edge (equivalently, we delete or add a single ``edge-hole"). This new edge or hole will form in the position that leads to the lowest total energy for that graph. Then, the new edge or hole is frozen in place and a new edge or hole is added. This approximation is useful, as finding the absolute ground state is equivalent to calculating the energy for $2^{N(N-1)/2}$ different graph configurations, which is not computationally plausible. Making use of graph isomorphism could greatly reduce the number of graphs one needs to check, but finding and identifying graph isomorphisms is a problem for which no P algorithm currently exists \cite{bib:GareyJohn1979}.

In \cite{bib:Wilk2014}, it was shown that when the model evolves epitaxially on 24 vertices from an empty initial graph, it achieves a local minimum 3-regular state which consists of four disjoint subgraphs, and a lower-energy 4-regular graph which consists of only 3 disjoint subgraphs. This indicates that the graph evolves from the empty starting point by forming small grains and merging them together, suggesting a process similar to the ripening that can be seen in many familiar material systems.

Consider a material system which is entirely in one phase, $\alpha$, which is suddenly brought into the region in its phase diagram of co-existence of two phases, $\alpha$ and $\beta$ (as may be achieved by a sudden change in temperature or pressure). The material will then be in a metastable state. Small regions of phase $\beta$ will begin to nucleate and form particles. If these particles are fairly diffuse and able to move free through phase $\alpha$, we may see the growth of the $\beta$ phase through a process called Ostwald Ripening \cite{bib:Ostwald}. Atoms at the surface of a particle are less tightly bound than atoms in the bulk, so they are more easily detached. Once detached, they diffuse through phase $\alpha$ until they re-attach to another particle of phase $\beta$. Since the surface-to-volume ratio is larger for small particles, these are more likely to lose atoms and less likely to gain them, so this leads to larger particles growing while smaller particles disappear. In this way, grains of phase $\beta$ form, until equilibrium between phases $\alpha$ and $\beta$ is achieved (or, if the relevant parameters are changed further, grains of $\beta$ will grow until they constitute the entire system).

The process of ripening in QG is quite different from that in material processes. The most striking difference is that Ostwald ripening occurs due to atoms being less tightly bound at surfaces than in the bulk of a material. Graphs in QG are non-oriented, so the usual intuition of a ``surface" does not exist. However, one may define a ``surface" in QG as a region of vertices which deviate from the ideal valence (or, away from the ground state, have a larger than average value of $|v_i - v_0|$). If one accepts this definition, then the difference in binding energy between vertices in the bulk at vertices at the surface is negligible. To be clear, neglecting the loops term in the Hamiltonian, the energy required to reduce the valence of a vertex in the bulk to zero is
\begin{equation}\
\frac{\Delta E_B}{g_V} = e^{pv_0^2} - e^{p(v_i - v_0)^2}
\end{equation}
where $v_i$ is the valence of the vertices within the bulk. The energy required to reduce the valence of a vertex in the surface to zero is
\begin{equation}
\frac{\Delta E_S}{g_V} = e^{pv_0^2} - e^{p(|v_i - v_0| + 1)^2}.
\end{equation}
On the other hand, the energy required to promote a vertex from the bulk to the surface of a grain is
\begin{equation}
\frac{\Delta E_{B\rightarrow S}}{g_V} = e^{p(|v_i - v_0| + 1)^2} - e^{p(v_i - v_0)^2}.
\end{equation}
To make this concrete, we calculate these quantities using the parameters of Konopka \textit{et al} \cite{bib:KMS2008} and taking the valence of the bulk vertices to be 2. This gives $\Delta E_B = 4.9017\times 10^4$, $\Delta E_S = 4.8899\times 10^4$ and $\Delta E_{B\rightarrow S} = 118.1903$. Here it can be seen explicitly that the energy difference between the bulk and the surface is negligible compared with the energy required to reduce the valence of either of them to zero.

The model will grow in way that reduces the number of vertices which have the maximum value of $|v_i - v_0|$, which means that when starting from the empty graph the first stage of evolution is to distribute disjoint edges throughout the graph. In the next stage of evolution some of these disjoint edges connect to form small ``grains". Eventually we have a graph which consists of many small disjoint subgraphs. It is from this stage that the process we call ripening can begin.

Consider the case of a $k$-regular graph consisting of several disjoint subgraphs where $k < v_0$. There are two possibilities for adding a single additional edge: the new edge will either connect two vertices already within the same grain, or it can connect two formerly disjoint grains to each other. Which one is favoured generally depends on the definition of loops we employ in $H_L$. Loop-counting based exclusively on shortest-path loops favours an edge which connects two formerly disjoint grains, as an internal edge within a grain will destroy some loops already in place. More general loop counting, in which all closed paths contribute to $H_L$ (as was used in \cite{bib:Wilk2014}), favours internal edge formation as this creates more loops, whereas connecting two separate grains does not (any walk that begins in one grain, crosses to the other and then crosses back again to the initial grain must traverse the newly formed edge twice, so such a walk cannot be a path and therefore cannot contribute even in this more general loop counting definition).

The situation from the complete graph is more difficult to interpret, as it is not easy to see which loops exist on the edges when looking only at the holes. In principle, evolution from the complete case should be able to give rise to grains of geometric space, as well as phase separation where some regions are in the pre-geometric state while others exhibit local emergent geometry. However, when the graph is already connected the different grains will be difficult to identify. Furthermore, algorithmic loop counting is computationally prohibitive when dealing with highly connected graphs.

\begin{figure}
	\centering
	\includegraphics[width=1\linewidth]{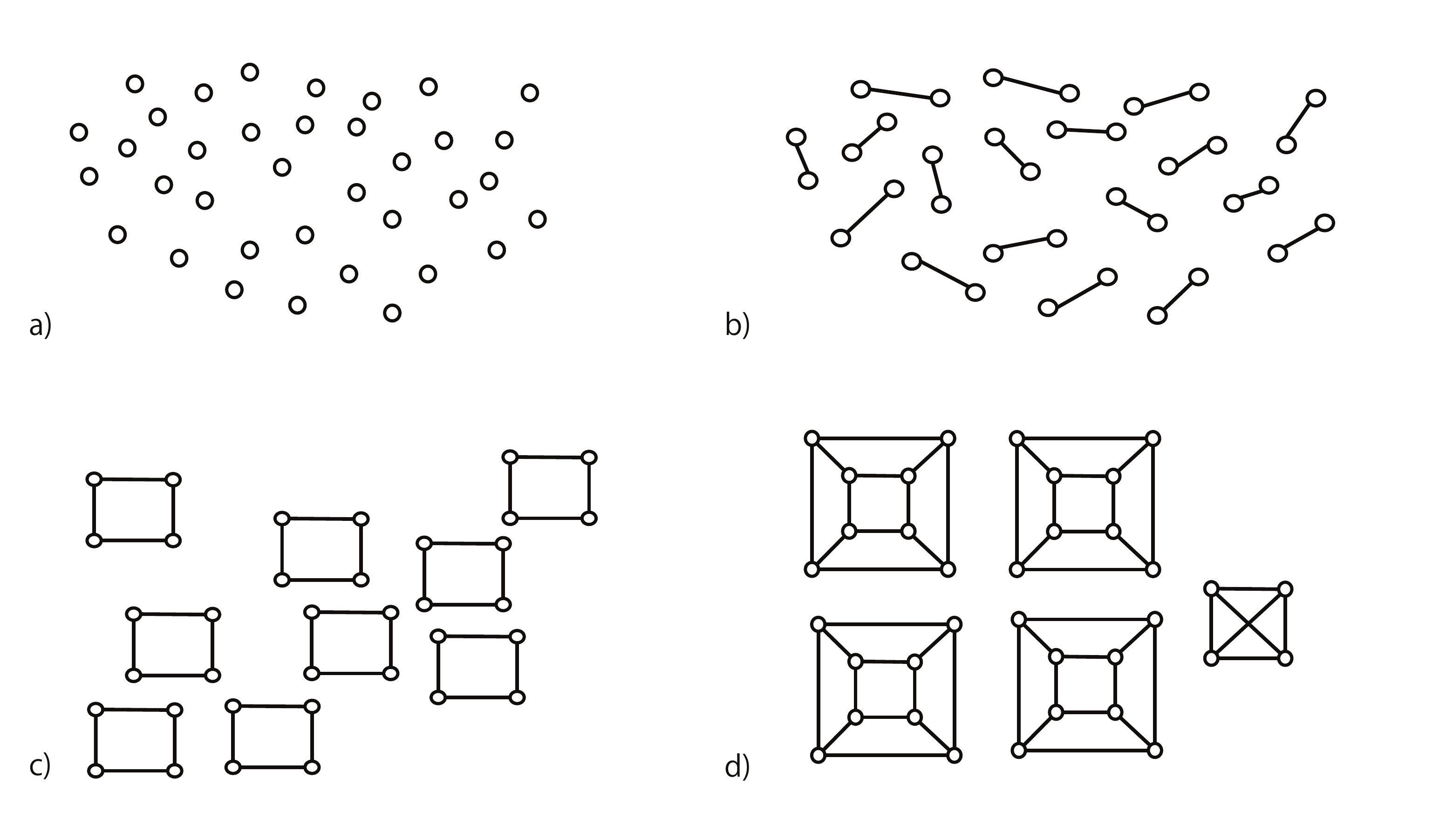}
	\caption{\label{fig:FromEmpty} a) The epitaxial growth of spatial domains starting with the empty graph as the initial graph. b) The graph always evolves in a way that reduces the number of vertices with the maximum value of $|v_i - v_0|$, so the first step of evolution is the formation of isolated edges. c) Isolated edges then form 4-loops, d) which knit together in a process we call ``ripening" to form cubic structures. The graph in d) is a local minimum, and is lower in energy than a connected, isotropic honeycomb lattice for the values of the parameters we have used.}
\end{figure}

The evolution from the complete graph to a low energy $v_0$-regular graph under the epitaxial approximation was simulated for $N=24$ and $N=36$ using parameters from Konopka \textit{et al} \cite{bib:KMS2008} ($g_V = 1$, $g_L = \dfrac{1}{500}$, $v_0 = 3$ and $p = 1.2$), the $r=6.5$ as discussed above. Important steps in the evolution of the $N=36$ are presented in Fig~\ref{fig:FromEmpty}. 

The general features of the evolution presented in Fig~\ref{fig:FromEmpty} are mostly the same as those that were seen in the $N=24$ case. The only significant difference is the fact that the $v_0$-regular state on $N=24$ consists entirely of the symmetrical cubic structures seen in Fig~\ref{fig:FromEmpty} d), whereas it is not possible to populate a graph on 36 vertices entirely with cubes as 36 is not divisible by 8, so a small $K_4$ forms from the ``leftover" vertices. Thus we have a frustrated ground state. Frustration is expected to be negligible in the $N \rightarrow \infty$ limit.

\begin{figure}
	\centering
	\includegraphics[width=1\linewidth]{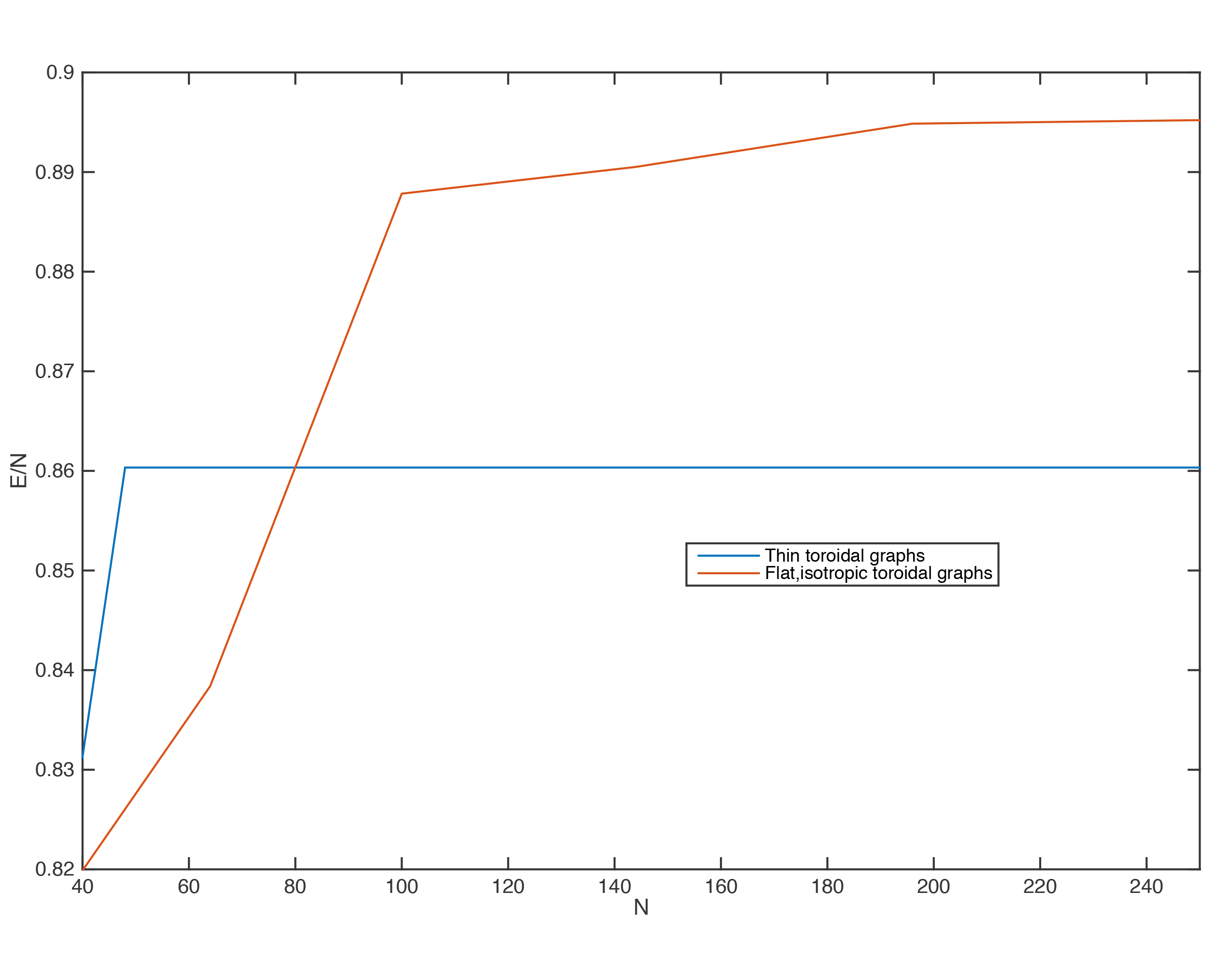}
	\caption{\label{fig:SubGraphs} Plot of the energy per vertex of various graphs as a function of the number of vertices $N$, using parameters based on Konopka \textit{et al} \cite{bib:KMS2008}. The red line shows the energy per vertex of isomorphic toroidal honeycomb graphs, where the major and minor circumferences are equal. The blue line shows honeycomb graphs where the minor circumference is fixed at 6, and only the major circumference is varied. For comparison, with our parameters a graph made up of isolated cubes always have an energy per vertex of 0.6385, which is notably lower than any of these honeycomb graphs. From this figure, the $N \rightarrow \infty$ limit is not certain, but there does not seem to be any indication that larger subgraphs will be more favourable at large $N$.}
\end{figure}

Somewhat surprisingly, the formation of isolated 4-loops in step c) of Fig~\ref{fig:FromEmpty} is more favourable than the formation of isolated 6-loops. This is because, although 6-loops reduce the energy more, the graph is able to produce more 4-loops, so the overall effect is a lower energy. Furthermore, epitaxial growth will favour the formation of 4-loops as these minimize the instantaneous energy at each step in evolution. Even more surprisingly, the ground state configuration in Fig~\ref{fig:FromEmpty} d) consists of isolated cubes, and not a connected honeycomb lattice. This cubic state is lower in energy than a connected, isotropic honeycomb lattice for the same values of the model parameters, the honeycomb having an energy of 29.4048 while the isolated cubes have an energy of 29.0779. This difference is small, but it is enough to demonstrate that the honeycomb graph is not the ground state, and the energy gap is expected to grow with $N$ (see Fig.~\ref{fig:SubGraphs} and the surrounding discussion). The honeycomb graph does contain more 6-loops than the isolated cubes (the honeycomb has $2/3$ loops per vertex, while the isolated cubes have $1/2$ in the ideal case and $2/9$ with the $K_4$ defect seen in Fig.~\ref{fig:FromEmpty} d), however the cubes also have many 4-loops. It is possible to increase the value of the Hamiltonian parameter $r$ to a value such that the 4-loops become negligible. For example, the honeycomb is lower in energy than the isolated cubes for $r = 7.1$ (the value used in previous work on QG \cite{bib:KMS2008}), however at this value 7-loops dominate over 6-loops. If a connected regular graph were to be favoured at all with this value, it would be a graph of heptagons, which would constitute a hyperbolic rather than a flat geometry. The crucial consequence of this discussion is that the connected honeycomb graph cannot be the ground state of the model, and this generic effect of favouring disconnected over connected graphs is likely to hold for other parameter choices.

\begin{figure}
	\centering
	\includegraphics[width=0.9\linewidth]{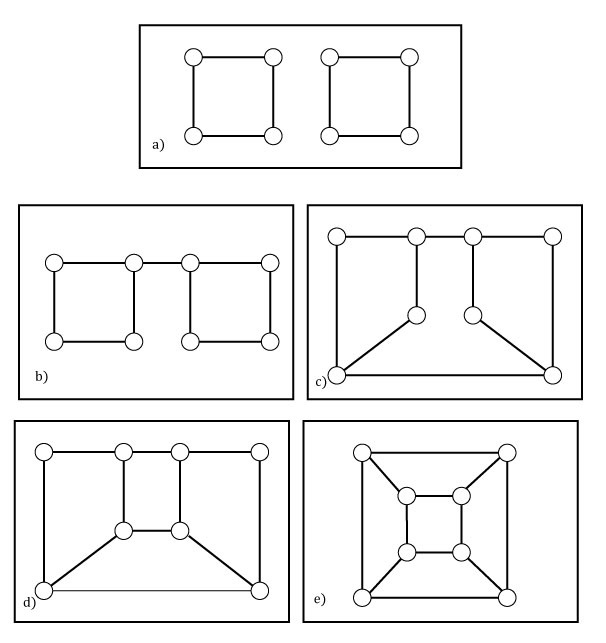}
	\caption{\label{fig:Ripe} The evolution of the graph from one consisting of 4-cycles seen in Fig~\ref{fig:FromEmpty} c) to one constituted by cubic structures seen in Fig~\ref{fig:FromEmpty} d) is presented step-by-step. Each subfigure in sequence represents the formation of one additional edge. Interestingly, the diameter of each grain increases from 2 in a) to 4 in b), before decreasing back to 3 in c). This is in contrast to the material version of Ostwald ripening, where the average diameter of large particles grows monotonically.}
\end{figure}

The process of ripening happens primarily between steps c) and d) in Fig~\ref{fig:FromEmpty}. Two ``squares" in c) knit together to form ``cubes". The cubes form one-by-one under the epitaxial approximation. The formation of a single cube from two squares is shown in Fig~\ref{fig:Ripe}.

In more traditional descriptions of ripening in material systems, one of the key quantities of interest is the average radius of particles as a function of time (as well as other factors such as the coefficient of diffusion). For abstract graphs with no inherent notion of geometry, the radius of the graph is defined as the smallest eccentricity of the graph, the diameter the largest eccentricity, where the eccentricity of a vertex is the largest possible geodesic distance to any other vertex on the graph. The case of ripening in QG as shown in Fig~\ref{fig:Ripe} is interesting because the diameter of the grains does not increase monotonically. Rather, we see the diameter of the grains undergo a sharp increase from 2 to 5 between steps a) and b), followed by a decrease to 3 in step c) (where it remains). This ``two steps forward one step back" type of growth is also seen when the graph transitions from one made up of isolated edges to one consisting of isolated 4-loops. The radius of the grains, however, does increase monotonically from 2 to 3. If the graph theoretic concepts of radius and diameter apply to the emergent geometry, this implies a picture where space forms by expansion, followed by slight contraction. However, the extent to which the graph theoretic notion of diameter of the grains corresponds to the geometric concept of the diameter of a space is unclear.

Another way in which the evolution observed here differs from ripening in material systems is that the process occurs mostly homogeneously. The 4-cycles in Fig~\ref{fig:FromEmpty} c) form one at a time, but cubic structures do not begin to form until there are no remaining vertices of degree 1. Likewise, the 4-cycles themselves do not begin to form until there are no remaining vertices of degree 0. The graph goes through distinct stages of evolution, but goes through them uniformly. This is in contrast to the nucleation and phase separation that is observed when material systems undergo ripening. This feature is most likely a consequence of both the epitaxial approximation, and the form of $H_V$, which causes evolution to focus on vertices of the maximum value of $|v_i - v_0|$ above all else.

Our simulation shows that the formation of two separate grains in this model is preferable to a regular lattice spreading out from the boundary of a honeycomb grain. We may then expect that a QG model with the empty graph as a high-energy starting point tends to form disjoint subgraphs. Such disconnected states can be lower in energy than a connected, extended lattice. Indeed, the cubic graph seen in Fig~\ref{fig:FromEmpty} d) is lower in energy than an isotropic connected honeycomb lattice on the same number of vertices. This undermines an assumption of the model that the Hamiltonian in equation \ref{eq:Ham} favours an extended honeycomb lattice.

Honeycomb lattice grains have the topology of a flat torus, so that there are no ``loose ends" (no vertices with $v_i \neq v_0$). This can be thought of as a rectangle with opposite edges identified, tiled with a honeycomb lattice. The ``lengths" of the two sides of this rectangle (alternatively, the minor and major circumference of the torus) are important, as when the sides are of length 6 then 6-loops can be formed by circumnavigating the space, not just by the plaquettes in the lattice. This increases the number of 6-loops and thereby decreases the energy. So, for a grain of a given $N$ which we assume to be connected, the lowest energy honeycomb configuration will be a torus which is long and thin, so that 6-loops can wrap around the width of the torus.

In figure \ref{fig:SubGraphs}, we show the energy of various graphs as a function of $N$, as calculated with algorithmic loop counting. The blue curve shows graphs which are flat tori with a minor circumference of 6, so that 6-loops may form by winding around the width of the torus. Therefore $N$ is changed by changing the major circumference, i.e. making the torus ``longer". The red curve shows the energy per vertex of isotropic toroidal graphs, where the major and minor circumferences are equal. It can be see that the 6-by-6 torus does indeed have the lowest energy per vertex of the honeycomb graphs shown, with the energy per vertex of the long thin tori becoming constant for large $N$. A graph made up of isolated cubes has an energy per vertex of 0.6385 with our parameters, which is significantly lower than any of the honeycomb graphs presented in Fig~\ref{fig:SubGraphs}. This shows a clear tendency for the formation of a disconnected space, which would make the model a poor description of our Universe. This is discussed further in the Conclusion.

It is worth noting that, although it is not always possible to tile an arbitrary space with a honeycomb lattice (see, for example, a buckyball, which must include some 5-loops in order to adopt spherical topology), this is a geometric constraint and has no bearing on the topology that a graph may adopt.

\section{Hypervalence}

The existence of a connected lattice-like ground state is an important requirement of QG. All of the terms in the Hamiltonian and all of the values used in our calculations were originally tailored to favour a connected honeycomb graph as the low energy state, however as we have seen there exist disjoint states of lower energy. The simplest solution to this problem is to modify the QG Hamiltonian in Eq~\ref{eq:Ham} by introducing a term which will drive connectivity. In particular, here we introduce a generalization of the valence term $H_V$ (Eq.~\ref{eq:Val}), which we call the ``hypervalence".

While the valence depends on the degree of each vertex, the hypervalence also depends on the 2-degree and 3-degree, \textit{i.e.} the number of vertices at distance 2 and 3 from the initial vertex. The hypervalence energy is defined
\begin{equation}
H_{HV} = \sum_i g_{v1}e^{p_1(v_{i,1} - v_1)^2} + g_{v2}e^{p_2(v_{i,2} - v_2)^2} + g_{v3}e^{p_3(v_{i,3}-v_3)^2}
\label{eq:Hypervalence}
\end{equation}
the parameters here are all analogous to those in the valence term. $v_{i,d}$ is the number of vertices at distance $d$ from vertex $i$, $g_{vd}$ is a positive real number determining the strength of the $d^{\textrm{th}}$ term and $p_d$ is a dimensionless real number determining the penalty for deviation from the ideal $d$-degree, $v_d$.

In principle, the number of terms within the hypervalence term could extend to $N$. Eq.~\ref{eq:Hypervalence} has been restricted to three terms for computational tractability, and therefore the weighting parameters $g_{vd}$ and $p_d$ must be selected so as to be negligible for $d>3$. Rapidly decaying couplings also implies a kind of pseudo-locality in the model - vertices an arbitrary distance away do affect the energy contribution from a given vertex, but this contribution quickly becomes negligible as the distance increases.

For the modified Hamiltonian to give rise to a regular ground state, we require $v_d = dv_1$. In keeping with the original model, we set $v_1 = v_0 = 3$, $g_{V1} = g_V = 1$ and $p_1 = p = 1.2$. For higher-degree terms to contribute negligibly, coupling constants were chose to scale as $g_{Vd} =  \left ( \frac{g_{V1}}{d!} \right )^d$ and the $p_d$ were fixed so that the argument of the exponential of each term in the hypervalence is equal for the empty graph. This leads to $p_3 = \frac{4p_2}{9} = \frac{p_1}{4}$. Other values are in principle possible, so long as higher-degree terms contribute significantly less than lower-degree terms.

\begin{figure}
	\centering
	\includegraphics[width=1\linewidth]{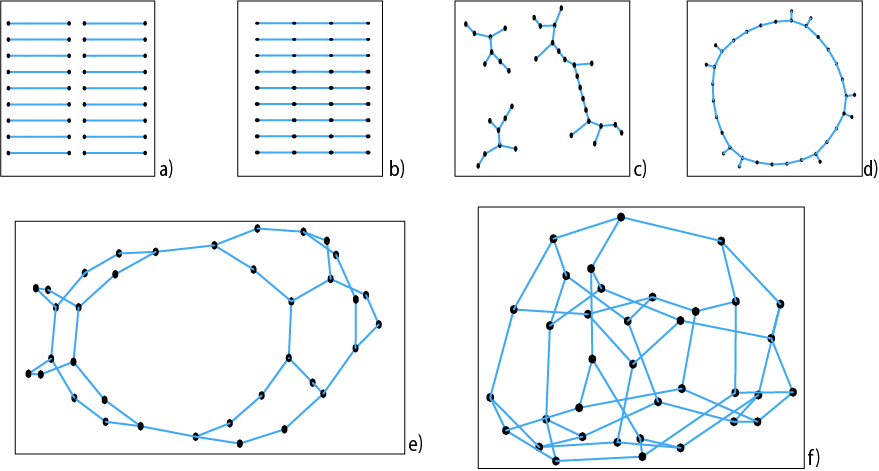}
	\caption{\label{fig:HypV} Stages of the ripening process under the hypervalence term. It can be seen that this does favour a connected graph, but the low energy state in f) is not lattice-like and there is no immediately evident emergent geometry. The lowest energy state here is still higher in energy than the regular 2-dimensional honeycomb lattice, indicating a failure of the epitaxial approximation to arrive at the ground state of the model.}
\end{figure}

Numerical simulation of the ripening of this new model was performed starting from an empty graph on $N=36$ under the epitaxial approximation, with various stages of the ripening process along with the final low-energy sate shown in Figure \ref{fig:HypV}. This low-energy graph is connected, making it a better representation of a continuous space than the low-energy graph in figure \ref{fig:Ripe}. However it is still not the kind of lattice-like graph hoped for by the QG program, nor does it have any obvious interpretation in terms of an emergent geometry.

Explicit calculations show, however, that with the hypervalence term, an isotropic connected honeycomb graph is lower in energy than both the isolated cubes and the low-energy state found by the epitaxial approximation shown in Fig.~\ref{fig:HypV}. The honeycomb graph has an energy of 38.5950, while the graph in Fig.~\ref{fig:HypV} f) has an energy of 46.2645 and the isolated cubes shown in Fig.~\ref{fig:FromEmpty} d) have an energy of $5.0830\times 10^4$. There are two important implications of this: 1) the addition of the hypervalence term is sufficient to favour an extended, flat, lattice-like ground state, and 2) the epitaxial approximation is not sufficient to find the ground state. The insufficiency of the simple epitaxial approximation employed here implies the importance of higher-order process, where multiple edges may be added or deleted simultaneously. However, simulating such processes is computationally difficult and beyond the scope of this work.

\section{Conclusion}

Ripening under the epitaxial approximation leads to the formation of a disconnected space. Clearly, a disconnected space is not a good representation of the Universe in which we live. The concept of disconnected space may hint at some interpretation in terms of a multiverse scheme, but only if any of the individual grains in the graph are large enough to support the large extended Universe we find ourselves in. This tendency to form disconnected spaces was noticed previously \cite{bib:Wilk2014}, and was a key motivation for this work.

The epitaxial approximation considered in this work only allowed for one-edge processes, where only one edge may be created or deleted at any time. Extending this approach to include the possibility of two-edge processes, where either two edges are created or deleted simultaneous, or where two edges hop to different positions simultaneously, may radically alter the graph dynamics presented here. It must be assumed that such processes are lower-probability events, but if they are sufficiently energetically favourable when compared with single-edge processes they may still dominate, or at least influence, graph evolution.

We have demonstrated that the honeycomb graph is not the ground state of the original QG model. Since the model was designed to give the honeycomb graph as a ground state, this may be considered a partial failure of the model. However, we have demonstrated that a simple extension of the model to include a hypervalence term restores the possibility that the connected honeycomb graph may be the ground state. Furthermore, we have shown that the epitaxial approximation employed in \cite{bib:Wilk2014} is not sufficient to return the true ground state of the model, at least when the model is extended to include a hypervalence term.

We have proposed a solution to this connectivity problem by including an additional term in the Hamiltonian, which we have called ``hypervalence". This term depends not only on the degree of each vertex but also on the 2- and 3-degree (and, ideally, all higher degrees in a rapidly converging manner). Hypervalence is a natural extension of the valence term in the original model, and its inclusion leads the model to favour a connected graph. While isolated cubes were found to have lower energy than a connected honeycomb lattice in the original model, the honeycomb lattice does become a lower energy state and a potential ground state when the hypervalence term is considered. However, the lowest energy state found under an epitaxial approach was not the honeycomb graph, nor any other lattice-like graph with a clear interpretation as a geometrical space. The lowest-energy state attained by this approach was also higher in energy than the honeycomb graph.

\section{Acknowledgements}
A. D. G. acknowledges the Australian Research Council for financial support (Contract No. DP130104381). We would like to thank Bill Moran for helpful comments and
conversations.

\end{document}